\documentclass[]{interact}

\usepackage{epstopdf}
\usepackage[caption=false]{subfig}

\usepackage[numbers,sort&compress]{natbib}
\bibpunct[, ]{[}{]}{,}{n}{,}{,}
\renewcommand\bibfont{\fontsize{10}{12}\selectfont}
\makeatletter
\def\NAT@def@citea{\def\@citea{\NAT@separator}}
\makeatother

\theoremstyle{plain}
\newtheorem{theorem}{Theorem}[section]

\newtheorem{proposition}[theorem]{Proposition}

\theoremstyle{definition}

\newtheorem{example}[theorem]{Example}

\theoremstyle{remark}
\newtheorem{remark}{Remark}

\usepackage{multirow}
\usepackage{color}

\begin{document}


\title{Controlling IER and EER in replicated regular two-level factorial experiments}

\author{
\name{Pengfei Li\textsuperscript{a}, Oludotun J. Akinlawon\textsuperscript{b} and Shengli Zhao\textsuperscript{c}\thanks{CONTACT Shengli Zhao. Email: zhaoshli758@126.com}}
\affil{\textsuperscript{a}Department of Statistics and Actuarial Science, University of Waterloo, Waterloo, ON, Canada, N2L 3G1;
\textsuperscript{b}Knowledge Utilization Studies Program, University of Alberta, Edmonton, AB, Canada, T6G 1C9;
\textsuperscript{c}School of Statistics, Qufu Normal University, Qufu, China, 273165}
}

\maketitle

\begin{abstract}
Replicated regular two-level factorial experiments are very useful for industry.
The goal of these experiments is to identify active effects that affect the mean and variance of the response.
Hypothesis testing procedures are widely used for this purpose.
However, the existing methods give results that are either too {anticonservative} or
conservative in controlling the individual and experimentwise error rates (IER and EER).
In this paper, we propose {a Monte Carlo method} and an exact-variance method to
identify active effects for the mean and variance, respectively, of the response.
Simulation studies show that our methods control the IER and
EER extremely well. Real data are used to illustrate the performance of the methods.
\end{abstract}


\section{Introduction}\label{secIntro}

Replicated regular two-level factorial experiments are widely used in industry \cite{Box78}.
In these experiments, several replications are available
for each treatment. The goal of these experiments is to
identify active effects that affect the mean and variance of the response. A
popular model for modeling the mean and variance of the response
simultaneously is the normal model with a linear regression for the mean and a
log-linear model for the variance \cite{Wu00, Wu09, Lou11}. The linear regression model
for the mean is called the location model, and the log-linear model for the
variance is called the dispersion model. Based on the above models, several
hypothesis-test procedures have been proposed to identify the active effects
for the mean and variance of the response \cite{Var05, Wu00, Wu09}.

To identify the active effects in the location model, Wu and Hamada \cite{Wu00, Wu09}
proposed a $t$-type statistic for each effect. To control the individual
error rate (IER), they suggested a $t$-distribution to calculate the critical
value, and to control the experimentwise error rate (EER), they suggested the studentized
maximum modulus distribution \cite{Wu00, Wu09}.
To identify the active effects in the dispersion
model, they proposed a $z$-type statistic for each
effect. They used the critical values calculated from the standard normal distribution
and the studentized maximum modulus distribution to control the IER and
EER, respectively. Variyath et al. \cite{Var05} suggested a jackknife method for the
replicated responses to provide an estimator of the variance of performance
measures such as the sample mean and the log of the sample variance of the replicated
responses at each treatment. They suggested an $F$ statistic
for each effect in the location and dispersion models. To control the IER,
they suggested using $F$ distributions to calculate the
critical values. However, they did not discuss how to control the EER.

Several methods have been proposed for the analysis of
unreplicated experiments in which no replication is available at
each treatment. These
methods can also be applied to replicated experiments. Some commonly used
methods include normal/half normal probability plots \cite{Dan59} and the
pseudo standard error estimation method proposed by Lenth \cite{Len89}. A detailed review of
methods for the analysis of unreplicated experiments is given by Hamada and
Balakrishnan \cite{Ham98}. Since these methods do not use information from
replications, they are expected to be less powerful than the
methods of Wu and Hamada \cite{Wu00, Wu09} and Variyath et al. \cite{Var05}.
Hereafter, we refer to the former as the
\textit{WH method} and to the latter as the \textit{VCA method}.

In Section \ref{simulation}, we investigate the performance of the WH and VCA
methods using simulation studies. The following is a summary of our findings.

\begin{itemize}
\item[(i)] Both methods can tightly control the IER in the location model when
the variances of the response are homogeneous across all the treatments, but they
inflate the IER when the variances of the response are heterogeneous.
The WH method has a similar problem with the EER.

\item[(ii)] Both methods are quite {anticonservative} for the control of the IER in the
dispersion model. The WH method is also {anticonservative} for the control of the EER in this model.
\end{itemize}

In this paper, we first identify the reasons for (i) and (ii). We then
propose {a Monte Carlo method} to calculate the critical values for the $t$-type
statistic in the WH method. The {Monte Carlo} method controls the
IER and EER well, whether the variances of the responses are homogeneous
or heterogeneous. We next suggest a new distribution for the $z$-type
statistic in the WH method and use it to derive the critical values for this
statistic. The $z$-type statistic coupled with the new critical
values controls the IER and EER well.

The paper is organized as follows. In Section \ref{secReview}, we review
the WH and VCA methods.
We further show that the two methods are equivalent in terms of identifying the
active effects in the location model. For comparison, we also review Lenth's
method. In Section \ref{our}, we  identify some reasons for findings (i) and
(ii) and present our new methods. Section
\ref{simulation} contains simulation studies comparing our methods with
the WH and VCA methods and Lenth's approach. In Section \ref{exam}, we apply all the
methods to a data set, and Section \ref{discussion} presents a summary {and discussion}.
{Some additional simulation studies and the R \cite{R} function for our methods
are provided in the supplementary material. }

\section{Review of Existing Methods}\label{secReview}

Let $y_{ij}$ be the response for the $i$th treatment and $j$th replication in
a two-level factorial experiment, $i=1,\ldots,m$ and $j=1,\ldots,n_{i}$. We
assume that the $y_{ij}$'s are independent. Suppose that there are $I$ effects of
interest. These effects can be main effects and interaction effects. Let
$x_{i1},\ldots,x_{iI}$ be the covariate values for these $I$ effects,
$i=1,\ldots,m$. For convenience of presentation, we consider the case where
$n_{1}=\cdots=n_{m}=n$.

The normal model described in Section \ref{secIntro} can be written as
\begin{equation}
y_{ij}\sim N(\mu_{i},\sigma_{i}^{2}),~i=1,\ldots,m,~j=1,\ldots,n \label{norm}%
\end{equation}
with
\begin{equation}
\mu_{i}=\alpha_{0}+\alpha_{1}x_{i1}+\ldots+\alpha_{I}x_{iI}%
~~\mbox{(location model)}, \label{loc}%
\end{equation}%
\begin{equation}
\log\sigma_{i}^{2}=\gamma_{0}+\gamma_{1}x_{i1}+\ldots+\gamma_{I}%
x_{iI}~~\mbox{(dispersion model)}. \label{dis}%
\end{equation}
To fit the above models, Wu and Hamada \cite{Wu00, Wu09} suggested first obtaining
the following summary statistics:
\[
\bar{y}_{i}=\sum_{j=1}^{n}y_{ij}/n~~\mbox{and}~~s_{i}^{2}=\sum_{j=1}^{n}%
(y_{ij}-\bar{y}_{i})^{2}/(n-1).
\]
We then regress $\bar{y}_{i}$ over $x_{i1},\ldots,x_{iI}$, $i=1,\ldots,m$, to
obtain the least square estimators of $\alpha_{0},\alpha_{1},\ldots,\alpha_{I}%
$, denoted by $\hat{\alpha}_{0},\hat{\alpha}_{1},\ldots,\hat{\alpha}_{I}$.
Next we regress $\log s_{i}^{2}$ over $x_{i1},\ldots,x_{iI}$, $i=1,\ldots,m$,
to obtain the least square estimators of $\gamma_{0},\gamma_{1},\ldots
,\gamma_{I}$, denoted by $\hat{\gamma}_{0},\hat{\gamma}_{1},\ldots,\hat
{\gamma}_{I}$.

Let
\[
\mbox{\bf x}_{l}=(x_{1l},\ldots,x_{ml})^{T}, l=1,\ldots,I,
\]
\[
\mbox{\bf z}_{1}=(\bar y_{1},\ldots,\bar y_{m})^{T},~ \mbox{
and
}~ \mbox{\bf z}_{2}=(\log s^{2}_{1},\ldots, \log s^{2}_{m})^{T}.
\]
For {balanced} two-level experiments, $\mbox{\bf x}_{l}$ has $m/2$ elements
equal to $-1$ and another $m/2$ equal to 1, and the $\mbox{\bf x}_{l}$'s are
mutually orthogonal. We then have
\begin{equation}
\label{alpha-gamma}\hat\alpha_{l}=\mbox{\bf x}_{l}^{T} \mbox{\bf z}_{1}/m ~
\mbox{and} ~ \hat\gamma_{l}=\mbox{\bf x}_{l}^{T} \mbox{\bf z}_{2}/m,~
l=1,\ldots,I.
\end{equation}

\subsection{WH Method}

\label{WH}

Wu and Hamada \cite{Wu00, Wu09} showed that under (\ref{norm})--(\ref{dis}),
\[
\hat\alpha_{l}\sim N\left(  \alpha_{l},\frac{1}{m^{2}n}\sum_{i=1}^{m}%
\sigma_{i}^{2} \right)  , ~l=1,\ldots,I.
\]
To test $H_{0}:\alpha_{l}=0$ in the location model, they
suggested a $t$-type statistic:
\[
t_{l}=\frac{\hat\alpha_{l}}{\sqrt{\frac{1}{m^{2}n}\sum_{i=1}^{m} s_{i}^{2}}},~
l=1,\ldots,I.
\]
To ensure the given level $\alpha$ for the IER for testing $H_0:\alpha_l=0$,
{which is
\begin{equation}
\label{IER.loc}
IER=Pr(\mbox{Rejecting }\alpha_l=0|H_0:\alpha_l=0),
\end{equation}}the critical value is set
to $t_{m(n-1),1-\alpha/2}$, i.e., the $(1-\alpha/2)$ upper quantile of a
$t$-distribution with $m(n-1)$ degrees of freedom. We reject $H_{0}:
\alpha_{l}=0$ if $|t_{l}|>t_{m(n-1),1-\alpha/2}$. To ensure the given level $\alpha$ for
the EER under the location model,
{which is
\begin{equation}
\label{EER.loc}
EER=Pr(\mbox{Rejecting at least one of }\alpha_l=0,~l=1,\ldots,I|H_0:\alpha_1=\cdots=\alpha_I=0),
\end{equation}}the critical value is set to $M_{I,m(n-1),1-\alpha}$,
the $(1-\alpha)$ quantile of a studentized maximum modulus distribution with
parameters $I$ and $m(n-1)$.

Under (\ref{norm})--(\ref{dis}), Wu and Hamada \cite{Wu00, Wu09} noted that
\[
(n-1)s_{i}^{2}=\sum_{j=1}^{n}(y_{ij}-\bar{y}_{i})^{2}\sim\sigma_{i}^{2}%
\chi_{n-1}^{2},
\]
where $\chi_{v}^{2}$ is the chi-squared distribution with $v$ degrees of
freedom. Taking the natural logarithm yields
\[
\log s_{i}^{2} \sim\log\sigma_{i}^{2}+\log\{\chi_{n-1}^{2}/(n-1)\}.
\]
Using the first-order Taylor expansion, they argued that we approximately have
\begin{equation}
\log s_{i}^{2}\sim N(\log\sigma_{i}^{2},{2}/{(n-1)}). \label{eq:app}%
\end{equation}

With
\begin{equation}
\label{var:gamma}E(\hat{\gamma}_{l})=\gamma_{l}, ~\mbox{Var}(\hat{\gamma}%
_{l})=\frac{1}{m^{2}}\sum_{i=1}^{m}\mbox{Var}(\log s_{i}^{2}),
\end{equation}
and (\ref{eq:app}), they found that $\hat{\gamma}_{l}$ has approximately
the distribution
\[
\hat{\gamma}_{l}\sim N(\gamma_{l},\frac{2}{m(n-1)}).
\]
They constructed a $z$-type test statistic
\[
z_{l}=\frac{\hat\gamma_{l}}{\sqrt{\frac{2}{m(n-1)}}}
\]
to test the null hypothesis $H_{0}:\gamma_{l}=0$.
To control the IER for testing $H_0:\gamma_l=0$,
{which is
\begin{equation}
\label{IER.disp}
IER=Pr(\mbox{Rejecting }\gamma_l=0|H_0:\gamma_l=0),
\end{equation}}they used the $N(0,1)$ distribution to calculate the critical value of the
$z$-type test statistic. To control the EER,
{which is
\begin{equation}
\label{EER.disp}
EER=Pr(\mbox{Rejecting at least one of }\gamma_l=0,~l=1,\ldots,I|H_0:\gamma_1=\cdots=\gamma_I=0),
\end{equation}}they used the studentized maximum
modulus distribution with parameters $I$ and $\infty$ to calculate the
critical value.

\subsection{VCA Method}

\label{VCA}

Variyath et al. \cite{Var05} suggested a jackknife method for the replicated
responses to provide an estimator of the variance of performance measures such
as $\bar y$ and $\log s^{2}$ of the replicated responses at each run. The
variance estimator of the performance measure is then used to estimate the
variance of the estimated factorial effects. Their method was applied to
control the IER only. To describe their method, we adopt their notation.
Let $y_{i}=(y_{i1},\ldots,y_{ij},\ldots,y_{in})$ be a random
sample of size $n$ for each run. Let $c(y_{i})$ be the performance measure of
interest. By deleting $y_{ij}$ from $y_{i}$ for $j=1,\ldots,n$, we obtain $n$
delete-one jackknife replicates of size $(n-1)$: $y_{i}({j}),$ $i=1$ $...,$
$m$. Hence, we obtain $n$ jackknife replications of the performance measure
$c(y_{i}({j}))$. The jackknife variance estimator of $c(y_{i})$ is
\[
\hat{V}_{ja}(c(y_{i}))=\frac{n-1}{n}\sum_{j=1}^{n}\Big(c(y_{i}({j}%
))-c(y_{i.})\Big)^{2},
\]
where $c(y_{i.})=\frac{1}{n}\sum_{j=1}^{n}c(y_{i}({j}))$. A pooled estimator of
the variance of $c(y_{i})$ is
\[
\hat{V}_{pja}(c(y))=\frac{1}{m}\sum_{i=1}^{m}\hat{V}_{ja}(c(y_{i})).
\]

Variyath et al. \cite{Var05} constructed an $F$-statistic
\[
F=\frac{\mbox{Mean~square~(MS)~for~the~factorial~effect}}{\hat{V}%
_{pja}(c(y))}
\]
to test the null hypothesis of interest. They showed theoretically
that the mean square of the factorial effect and $\hat
{V}_{pja}(c(y))$ are independent. They used the $F$-distribution with degrees of freedom
$1$ and $m(n-1)$ to calculate the critical value of the above $F$-statistic.

For the location model, $c(y_{i})=\bar y_{i}$ and $\hat{V}_{pja}%
(c(y))=\frac{1}{m}\sum_{i=1}^{m}\hat{V}_{ja}(\bar y_{i})$. The $F$-statistic
to test the hypothesis $H_{0}:\alpha_{l}=0$ can be written as
\begin{equation}
\label{eq:F}F_{l}=\frac{\mbox{MS}(\hat{\alpha}_{l})}{\frac{1}{m}\sum_{i=1}%
^{m}\hat{V}_{ja}(\bar y_{i})},
\end{equation}
where $\mbox{MS}(\hat{\alpha}_{l})=\hat\alpha_{l}^{2}(\mathbf{x}_{l}%
^{T}\mathbf{x}_{l})=m\hat\alpha_{l}^{2}$.

For the dispersion model, $c(y_{i})=\log s_{i}^{2}$ and $\hat{V}%
_{pja}(c(y))=\frac{1}{m}\sum_{i=1}^{m}\hat{V}_{ja}(\log s_{i}^{2})$. The
$F$-statistic to test the hypothesis $H_{0}:\gamma_{l}=0$ is
\[
F_{l}=\frac{\mbox{MS}(\hat{\gamma}_{l})}{\frac{1}{m}\sum_{i=1}^{m}\hat{V}%
_{ja}(\log s_{i}^{2})},
\]
where $\mbox{MS}(\hat{\gamma}%
_{l})=\hat\gamma_{l}^{2}(\mathbf{x}_{l}^{T}\mathbf{x}_{l})=m\hat\gamma_{l}%
^{2}$. The simulation results of Variyath et al. \cite{Var05} show that the jackknife variance estimator works
well for $\log s_{i}^{2}$ when $n\geq50$. For small $n$, they considered an
adjustment factor for the variance estimator of $\log s_{i}^{2}$.

\subsection{Connection Between WH and VCA Methods}

\label{connection}

The WH and VCA methods can both be applied to control the IER in the
location model. In this subsection, we present a proposition showing
the connection between these methods.

\begin{proposition}
For testing $H_{0}:\alpha_{l}=0$, we have
\[
t_{l}^{2}=F_{l},
\]
where $F_{l}$ is defined in (\ref{eq:F}). Therefore, the two methods are
equivalent in terms of controlling the IER.
\end{proposition}

\noindent\textit{\textbf{Proof.}} Recall that the jackknife variance estimator
of the performance measure of interest, $c(y_{i})$, is
\[
\hat{V}_{ja}(c(y_{i}))=\frac{n-1}{n}\sum_{j=1}^{n}\Big(c(y_{i}({j}%
))-c(y_{i.})\Big)^{2},
\]
where $c(y_{i.})=\frac{1}{n}\sum_{j=1}^{n}c(y_{i}({j}))$.

For the location model, we have $c(y_{i})=\bar{y}_{i}$,
\[
c(y_{i}({j})) = \frac{\sum_{k\neq j}y_{ik}}{n-1}=\frac{\sum_{k=1}^{n}%
y_{ik}-y_{ij}}{n-1} = \frac{n\bar y_{i}-y_{ij}}{n-1},
\]
and
\[
c(y_{i.}) = \frac{1}{n}\sum_{j=1}^{n}c(y_{i}({j})) =\frac{1}{n}\sum_{j=1}%
^{n}\left(  \frac{n\bar y_{i}-y_{ij}}{n-1}\right)  =\bar y_{i}.
\]
Therefore,
\[
\hat{V}_{ja}(c(y_{i})) = \frac{n-1}{n}\sum_{j=1}^{n} \left(  \frac{n\bar
y_{i}-y_{ij}}{n-1}-\bar y_{i} \right)  ^{2} = \frac{n-1}{n}\sum_{j=1}^{n}
\left(  \frac{\bar y_{i}-y_{ij}}{n-1} \right)  ^{2} = \frac{\sum_{j=1}%
^{n}\left(  y_{ij}-\bar{y}_{i}\right)  ^{2}}{(n-1)n}.
\]
Thus, the jackknife variance estimator for $c(y_{i})$ becomes
\[
\hat{V}_{ja}(\bar{y}_{i})=\frac{s_{i}^{2}}{n}.
\]
Then, a pooled estimator of $c(y_{i})$ is
\[
\hat{V}_{pja}(c(y))=\frac{1}{m}\sum_{i=1}^{m}\hat{V}_{ja}(c(y_{i}))=\frac
{1}{mn}\sum_{i=1}^{m}s_{i}^{2}.
\]
Therefore, the $F$-statistic to test the hypothesis $H_{0}:\alpha_{l}=0$ can
be written as
$$
\label{eq:variy}F_{l}=\frac{\mbox{MS}(\hat{\alpha}_{l})}{\frac{1}{m}\sum
_{i=1}^{m}\hat{V}_{ja}(\bar y_{i})}=\frac{\hat{\alpha}_{l}^{2}m}{\frac{1}%
{mn}\sum_{i=1}^{m}s_{i}^{2}}=t_{l}^{2}.
$$
This completes the proof.

\subsection{Lenth's Method}

\label{Lenth}

Lenth \cite{Len89} proposed a robust estimator of the standard deviation of the
factorial effects of interest. His approach is the same for both the
dispersion and location models; we describe the method for the dispersion model
only. Suppose that $\hat{\gamma}_{1},\ldots,\hat{\gamma}_{I}$ are the least square
estimators of the factorial effects ($\gamma_{1},\ldots,\gamma_{I}$) in
the dispersion model.

Lenth \cite{Len89} proposed a pseudo standard error (PSE) for the standard deviation
of $\hat\gamma_{l}$:
$$
\label{eq:pse}\mbox{PSE}=1.5\cdot\mbox{median}_{\{_{|\hat{\gamma}%
_{l}|<2.5s_{0}}\}}|\hat{\gamma}_{l}|.
$$
Here the median is computed among the $|\hat{\gamma}_{l}|$'s with
$|\hat{\gamma}_{l}|<2.5s_{0}$ and $s_{0}=1.5\cdot\mbox{median}|\hat{\gamma
}_{l}|.$ He defined a $t$-type statistic
\[
t_{\mbox{{\scriptsize Lenth}},l}=\frac{\hat{\gamma}_{l}}{PSE}
\]
to test the hypothesis $H_{0}:\gamma_{l}=0$.

Lenth's method does not require an unbiased estimator of the variance of the response.
For this reason, researchers have used his method for both replicated and
unreplicated factorial experiments. The critical values for controlling the IER
and EER are given in Appendix H of Wu and Hamada \cite{Wu00, Wu09}.

\section{Our Methods}

\label{our}

\subsection{Location Model}

\label{location}

Recall that the $t$-type test statistic for testing $H_{0}:\alpha_{l}=0$ in
the location model is
\[
t_{l}=\frac{\hat\alpha_{l}}{\sqrt{\frac{1}{m^{2}n}\sum_{i=1}^{m} s_{i}^{2}}},
\]
and the variance of $\hat\alpha_{l}$ is $\mbox{Var}(\hat\alpha_{l})=\frac
{1}{m^{2}n}\sum_{i=1}^{m}\sigma_{i}^{2}$. Next we try to find the distribution
of $t_{l}$ under the null hypothesis.

Note that the $t$-type test statistic can be rewritten as
\begin{equation}
t_{l}=\frac{{\hat{\alpha}_{l}}\Big/\sqrt{\frac{1}{m^{2}n}\sum_{i=1}^{m}%
\sigma_{i}^{2}}}{{{\sqrt{\frac{1}{m^{2}n}\sum_{i=1}^{m}s_{i}^{2}}%
\Big/\sqrt{\frac{1}{m^{2}n}\sum_{i=1}^{m}\sigma_{i}^{2}}}}}. \label{sec2.2.1}%
\end{equation}
The classical theory of the normal distribution implies that the numerator of
(\ref{sec2.2.1}) is independent of the denominator. Further, under the null
hypothesis, the numerator has the $N(0,1)$ distribution, i.e.,
\[
{\hat{\alpha}_{l}}\left/  \sqrt{(m^{2}n)^{-1}\sum_{i=1}^{m}\sigma_{i}^{2}}\sim
N(0,1)\right.  .
\]
The denominator of (\ref{sec2.2.1}) can be expressed as
\[
\frac{\sum_{i=1}^{m}s_{i}^{2}}{\sum_{i=1}^{m}\sigma_{i}^{2}}=\frac{1}{n-1}%
\sum_{i=1}^{m}\left[  \frac{(n-1)s_{i}^{2}}{\sigma_{i}^{2}}\left(
\frac{\sigma_{i}^{2}}{\sum_{i=1}^{m}\sigma_{i}^{2}}\right)  \right]  .
\]
Let
\[
\rho_{i}^{2}=\frac{\sigma_{i}^{2}}{\sum_{i=1}^{m}\sigma_{i}^{2}}%
.\label{eq:1.11}%
\]
Note that for $i=1,\ldots,m,$
\[
\frac{(n-1)s_{i}^{2}}{\sigma_{i}^{2}}\sim\chi_{n-1}^{2}%
\]
and they are independent. Therefore, the denominator of (\ref{sec2.2.1})
follows a weighted sum of $m$ independent $\chi_{n-1}^{2}$ distributions. For
convenience of presentation, we write
\begin{equation}
\frac{\sum_{i=1}^{m}s_{i}^{2}}{\sum_{i=1}^{m}\sigma_{i}^{2}}\sim\sum_{i=1}%
^{m}\rho_{i}^{2}\chi_{n-1}^{2}/(n-1). \label{denom.dis}%
\end{equation}
Hence, under the null hypothesis of $\alpha_{l}=0,$ the distribution of the
$t$-type statistic of (\ref{sec2.2.1}) is
\begin{equation}
t_{l}\sim\frac{N(0,1)}{\sqrt{\sum_{i=1}^{m}\rho_{i}^{2}\chi_{n-1}^{2}/(n-1)}}.
\label{tl.dis}%
\end{equation}
In this expression, the $N(0,1)$ random variable and the $m$ $\chi
_{n-1}^{2}$ random variables are independent.

\begin{remark}{1.}
If the $\sigma_{i}^{2}$'s are homogeneous, then
\[
\frac{\sigma_{i}^{2}}{\sum_{i=1}^{m}\sigma_{i}^{2}}=\frac{\sigma^{2}}%
{m\sigma^{2}}=\frac{1}{m}.
\]
Therefore,
\[
\frac{\sum_{i=1}^{m}s_{i}^{2}}{\sum_{i=1}^{m}\sigma_{i}^{2}}\sim\frac
{\chi_{m(n-1)}^{2}}{m(n-1)}.
\]
Thus, under $H_{0}:\alpha_{l}=0$, we obtain
\[
t_{l}\sim\frac{N(0,1)}{\sqrt{\frac{\chi_{m(n-1)}^{2}}{m(n-1)}}}=t_{m(n-1)}.
\]
However, if the $\sigma_{i}^{2}$'s are heterogeneous, then the distribution of
$t_{l}$ may not be a $t$-distribution under the null hypothesis.
\end{remark}

We now consider how to control the IER defined in (\ref{IER.loc}) for the location model.
The explicit form of
the cumulative distribution function of $t_{l}$ in (\ref{tl.dis}) is unknown
if the $\sigma_{i}^{2}$'s are heterogeneous. However, this suggests a way to
generate random samples from this distribution, which can be used to
calculate the critical value for controlling the IER. We
propose below {a Monte Carlo method} to generate random samples from the
distribution in (\ref{tl.dis}). Since the $\rho_{i}^{2}$'s are unknown, we
estimate them from the given data via $\hat{\rho}_{i}^{2}=s_{i}^{2}/\sum
_{k=1}^{m}s_{k}^{2}$.

\begin{itemize}
\item[] \textbf{Step 1}: Compute $\hat\rho_{i}^{2}$, for $i=1,2,\dots,m$, from
the given data set.

\item[] \textbf{Step 2}: For $b=1,\ldots,M$,

\begin{itemize}
\item[] \textbf{Step 2.1}: Generate a $N(0,1)$ random variable $U_{b}$.

\item[] \textbf{Step 2.2}: Generate $m$ independent $\chi_{n-1}^{2}$ random
variables $V_{b1},\dots,V_{bm}$.

\item[] \textbf{Step 2.3}: Compute $t^{(b)}=\frac{U_{b}}{\sqrt{{\sum_{i=1}%
^{m}V_{bi}\hat\rho_{i}^{2}}/{(n-1)}}}$.
\end{itemize}

\item[] \textbf{Step 3}: The critical value $C_{IER}$ for controlling IER in the
location model at the given $\alpha$ value is set to the $(1-\alpha/2)$ upper
quantile of $\{t^{(b)}, b=1,2,...,M\}$.
\end{itemize}

In R, the critical
value $C_{IER}$ can be calculated rapidly. With our R function in the supplementary material, it takes just seconds to get the
$C_{IER}$ for $M=100,000$.

We then consider controlling the EER in (\ref{EER.loc}) for the location model, i.e.,
\[
\mbox{EER}=Pr(\max_{l}|t_{l}|\geq C_{EER}|H_{0}:\alpha_{1}=\cdots=\alpha
_{I}=0)
\]
at a given $\alpha$ level. We first use the result in (\ref{sec2.2.1}) to
investigate the distribution of $\max_{1\leq l\leq I}|t_{l}|$. Note that
\begin{align}
\mbox{max}_{1\leq l\leq I}|t_{l}|  &  =\mbox{max}_{1\leq l\leq I}%
\frac{\left\vert {\hat{\alpha}_{l}}\Big/\sqrt{\frac{1}{m^{2}n}\sum_{i=1}%
^{m}\sigma_{i}^{2}}\right\vert }{{\sqrt{\frac{1}{m^{2}n}\sum_{i=1}^{m}%
s_{i}^{2}}\Big/\sqrt{\frac{1}{m^{2}n}\sum_{i=1}^{m}\sigma_{i}^{2}}}%
}\nonumber\label{sec2.2.31}\\
&  =\frac{\mbox{max}_{1\leq l\leq I}\left\vert {\hat{\alpha}_{l}}%
\Big/\sqrt{\frac{1}{m^{2}n}\sum_{i=1}^{m}\sigma_{i}^{2}}\right\vert }%
{{\sqrt{\frac{1}{m^{2}n}\sum_{i=1}^{m}s_{i}^{2}}\Big/\sqrt{\frac{1}{m^{2}%
n}\sum_{i=1}^{m}\sigma_{i}^{2}}}}.
\end{align}
The distribution of the denominator of (\ref{sec2.2.31}) has been investigated
in (\ref{denom.dis}). We now study the distribution of the numerator. Note
that
\[
(\hat{\alpha}_{1},\cdots,\hat{\alpha}_{I})^{T}=\frac{1}{m}\mathbf{X}%
^{T}\mbox{\bf z}_{1}%
\]
with {\bf$\mathbf{X}=(\mbox{\bf x}_{1},\ldots,\mbox{\bf x}_{I})$.} Under the normal
assumption in (\ref{norm}) on $y_{ij}$, $\mbox{\bf z}_{1}$
follows a multivariate normal distribution (MVN). By the properties of
the MVN, $(\hat{\alpha}_{1}%
,\cdots,\hat{\alpha}_{I})^{T}$ is multivariate normally distributed. Under the
null hypothesis $H_{0}:\alpha_{1}=\cdots=\alpha_{I}=0$,
\[
(\hat{\alpha}_{1},\cdots,\hat{\alpha}_{I})^{T}\sim\mbox{MVN}\left(\mathbf{0}%
,\frac{1}{m^{2}}\mathbf{X}^{T}\mbox{Var}(\mbox{\bf z}_{1})\mathbf{X}\right).
\]
Using the fact that $\mbox{Var}(\mbox{\bf z}_{1})=\mbox{diag}\{\sigma_{1}%
^{2}/n,\ldots,\sigma_{m}^{2}/n\}$, we have
\[
\left(  \frac{\hat{\alpha}_{1}}{\sqrt{\frac{1}{m^{2}n}\sum_{i=1}^{m}\sigma
_{i}^{2}}},\ldots,\frac{\hat{\alpha}_{I}}{\sqrt{\frac{1}{m^{2}n}\sum_{i=1}%
^{m}\sigma_{i}^{2}}}\right)  ^{T}\sim\mbox{MVN}(\mathbf{0},\mathbf{X}%
^{T}\mbox{diag}\{\rho_{1}^{2},\ldots,\rho_{m}^{2}\}\mathbf{X}).
\]

Combining the distributions of the numerator and denominator of
(\ref{sec2.2.31}), we see that $\max_{1\leq l\leq I}|t_{l}|$ has the same
distribution as the ratio {\bf$U/\sqrt{V}$} such that

\begin{itemize}
\item[(1)] $U$ and $V$ are independent;

\item[(2)] $U$ has the same distribution as the maximum of the absolute values
of an $I$-dimensional multivariate normal random vector with mean vector 0 and
variance-covariance matrix $\mathbf{X}^{T}\mbox{diag}\{\rho_{1}^{2}%
,\ldots,\rho_{m}^{2}\}\mathbf{X}$;

\item[(3)] The distribution of $V$ is a weighted sum of $m$ independent
$\chi_{n-1}^{2}$ distributions, i.e., $V\sim\sum_{i=1}^{m}\rho_{i}^{2}%
\chi_{n-1}^{2}/(n-1)$.
\end{itemize}

The explicit form of the cumulative distribution function of $\max_{1\leq
l\leq I}|t_{l}|$ may be unknown. However, this suggests a way to generate a
random sample from the distribution as follows:

\begin{itemize}
\item[] \textbf{Step 1}: Compute $\hat\rho_{i}^{2}$, for $i=1,2,\dots,m$, from
the given data set.

\item[] \textbf{Step 2}: For $b=1,\ldots,M$,

\begin{itemize}
\item[] \textbf{Step 2.1:} Generate an $I$-dimensional random vector
$(U_{b1},\cdots,U_{bI})^{T}$ from the multivariate normal distribution with
mean vector 0 and variance-covariance matrix $\mathbf{X}^{T}\mbox{diag}\{\hat
{\rho}_{1}^{2},\ldots,\hat{\rho}_{m}^{2}\}\mathbf{X}$.

\item[] \textbf{Step 2.2}: Generate $m$ independent $\chi_{n-1}^{2}$ random
variables $V_{b1},\dots,V_{bm}$.

\item[] \textbf{Step 2.3}: Compute $t^{(b)}_{l}=\frac{U_{bl}}{\sqrt{{\sum
_{i=1}^{m}V_{bi}\hat\rho_{i}^{2}}/{(n-1)}}}$, $l=1,\ldots,I$.

\item[] \textbf{Step 2.4}: Compute $\max_{1\leq l\leq I}| t_{l}^{(b)} |$.
\end{itemize}

\item[] \textbf{Step 3}: The critical value $C_{EER}$ for controlling EER in the
location model at the given $\alpha$ value is set to the $(1-\alpha)$ upper
quantile of $\{\max_{1\leq l\leq I}| t_{l}^{(b)} |, b=1,2,...,M\}$.
\end{itemize}
\vspace{-0.1in}

We make two remarks. First, we can show that if the $\sigma_{i}^{2}$'s are
homogeneous then $\max_{1\leq l\leq I}|t_{l}|$ follows a studentized maximum
modulus distribution with the parameters $I$ and $m(n-1)$; this distribution is suggested
by Wu and Hamada \cite{Wu00, Wu09} to control the EER in the location model. However,
if the $\sigma_{i}^{2}$'s are heterogeneous, $\max_{1\leq l\leq I}|t_{l}|$ no
longer follows this distribution. Our method can be applied to both
situations.
{Second,  the distributions of $U$ and $V$  are known in the literature.
The random variable $U$ has the same distribution as the maximum of $I$ correlated folded normal random variables,
and the distribution of $V$ is a weighted sum of $\chi^2$-distributions \cite{Sol77, Gab87, Cas05}.
In Step 2 above, we directly generate two independent random variables $U^{(l)}$ and $V^{(l)}$ from the  distributions of $U$ and $V$, respectively. Then $U^{(l)}/\sqrt{V^{(l)}}$ is a random variable from the distribution of {\bf$U/\sqrt{V}$}. This method is equivalent to the one outlined in Steps 2.1 to 2.4 above. Since the idea of our method is quite natural,
we implement it in the R function in the supplementary material.
The computational cost is again low. Our R function
takes just seconds to obtain $C_{EER}$ when $M=100,000$.}

\subsection{Dispersion Model}

\label{dispersion}


Recall that for testing $H_{0}:\gamma_{l}=0$ in the dispersion model, the
$z$-type statistic is defined as
\[
z_{l}=\frac{\hat\gamma_{l}}{\sqrt{\frac{2}{m(n-1)}}}.
\]
The actual variance of $\hat\gamma_{l}$ is given in (\ref{var:gamma}) as
\[
\mbox{Var}(\hat\gamma_{l})=\frac{1}{m^{2}}\sum_{i=1}^{m}\mbox{Var}(\log
s_{i}^{2}).
\]

For the small-$n$ case, instead of using the approximation $2/(n-1)$ to
$\mbox{Var}(\log s_{i}^{2})$, we suggest using the exact variance of the $\log
s_{i}^{2}$. Note that
\[
\log s_{i}^{2}\sim\log\sigma_{i}^{2}+\log(\chi_{n-1}^{2}/(n-1)).
\]
Therefore,
\[
\mbox{Var}(\log s_{i}^{2})=\mbox{Var}\left(  \log\frac{\chi_{n-1}^{2}}%
{n-1}\right)  =\mbox{Var}(\log\chi_{n-1}^{2})
\]
and
\[
\mbox{Var}(\hat{\gamma}_{l})=\frac{1}{m}\mbox{Var}(\log\chi_{n-1}^{2}).
\]
Here $\mbox{Var}(\log\chi_{n-1}^{2})$ means the variance of the logarithm of a
random variable from the $\chi_{n-1}^{2}$ distribution.

Let
\[
a_{n}=\sqrt{ \frac{\mbox{Var}(\log{\chi_{n-1}^{2}} )}{2/(n-1)} },
\]
which is the square root of the ratio of the true variance over the approximate
variance for $\log s_{i}^{2}$. The values of $a_{n}$ for some small $n$ values
are given in Table 1. \vspace{0.1in}

{\renewcommand{\arraystretch}{1}
\begin{table}[!th]
\caption{Comparison of the exact and approximate variances of $\log s_{i}^{2}$.}
\label{table1}
\vspace{0.1in}
\centering
{\tabcolsep=2.5mm
\begin{tabular}[c]{ccccccccc}\hline
\multirow{2}*{Variance} & \multicolumn{8}{c}{$n$}\\
\cline{2-9}
& 3 & 4 & 5 & 6 & 7 & 8 & 9 & 10\\
\hline
Exact & 1.645 & 0.935 & 0.645 & 0.490 & 0.395 & 0.330 & 0.284 & 0.249\\
Approximate & 1.000 & 0.667 & 0.500 & 0.400 & 0.333 & 0.286 & 0.250 & 0.222\\
$a_{n}$ & 1.283 & 1.184 & 1.136 & 1.107 & 1.088 & 1.075 & 1.066 & 1.058\\
\hline
\end{tabular}
}
\end{table}
}

We observe that the $z$-type test statistic can be written as
\[
z_{l}=\frac{\hat\gamma_{l}} { \sqrt{\frac{1}{m}\mbox{Var}(\log{\chi_{n-1}^{2}%
})} } \sqrt{ \frac{\mbox{Var}(\log{\chi_{n-1}^{2}})}{2/(n-1)} } =a_{n}
\frac{\hat\gamma_{l}} {\sqrt{\mbox{Var}(\hat\gamma_{l})}}.
\]
Motivated by this form of the statistic, we suggest using
$N(0,a_{n}^{2})$ to approximate the true distribution of the $z$-type
statistic under the null hypothesis $H_{0}: \gamma_{l}=0$.

We observe that the true distribution of $\hat{\gamma}_{l}$ may not be normal.
However, since $\hat{\gamma}_{l}$ as given in (\ref{alpha-gamma}) is a linear
combination of $m$ independent and identically distributed log-transformed random
variables, the central limit theorem implies that the distribution of
$\hat{\gamma}_{l}$ may be well approximated by the normal distribution.
Simulation studies show that the normal approximation works well even for
small $n$ and $m$, e.g., $m=8$ and $n=3$.

From Table 1, our suggested distribution $N(0,a_{n}^{2})$ and the $N(0,1)$
distribution suggested by Wu and Hamada \cite{Wu00, Wu09} can be quite different
for small $n$. The two distributions become similar as $n$ becomes
large. However, our distribution based on the exact variance of
$\log s_{i}^{2}$ may be preferable for small $n$ in practical applications.
This has been verified in our simulation study.


With the suggested distribution $N(0,a_{n}^{2})$, to control
the IER in (\ref{IER.disp}) for the dispersion model
at the given $\alpha$ level, we can set the critical value $C_{IER}$
to be the upper $(1-\alpha/2)$ quantile of the $N(0,a_{n}^{2})$. That is,
\[
C_{IER}=a_{n}\Phi^{-1}(1-\alpha/2)
\]
where $\Phi(\cdot)$ is the cumulative distribution function of $N(0,1)$.

To control the EER in (\ref{EER.disp}) for the dispersion model, we note that
\begin{align*}
EER  &  =Pr\left(  \max_{1\leq l\leq I}|z_{l}|\geq C_{EER}|H_{0}:\gamma
_{1}=\cdots=\gamma_{I}=0\right) \\
&  =1-Pr\left(  \max_{1\leq l\leq I}|z_{l}/a_{n}|<C_{EER}/a_{n}|H_{0}%
:\gamma_{1}=\cdots=\gamma_{I}=0\right) \\
&  =1-\{\Phi(C_{EER}/a_{n})-\Phi(-C_{EER}/a_{n})\}^{I}\\
&  =1-\{2\Phi(C_{EER}/a_{n})-1\}^{I}.
\end{align*}
Here $C_{EER}$ is the critical value for controlling the EER. Therefore,
to control the EER at the given $\alpha$ level
\[
C_{EER}=a_{n}\Phi^{-1}\left(  0.5+0.5(1-\alpha)^{1/I}\right).
\]

\section{Simulation Study}

\label{simulation}

In this section, we perform simulations to compare the performance
of our methods for controlling the IER and EER with the three existing
methods for both location and dispersion models.

\subsection{Simulation Results for Location Model}

We first present the results for controlling the IER. As discussed in Section
\ref{connection}, the WH and VCA methods are equivalent for controlling the IER
in the location model. Therefore, we compare only our
method, the WH method, and Lenth's method.
We consider two cases: homogenous $\sigma_{i}^{2}$'s and
heterogeneous $\sigma_{i}^{2}$'s.

\noindent\textbf{Case I: }$\sigma_{i}^{2}$ \textbf{homogeneous}

In this case, we perform simulations of $2^{3}$ and $2^{4}$ factorial
experiments. For the $2^{3}$ experiment with three two-level factors, $A$, $B$,
and $C$, we use the model
\[
y_{ij}\sim N(10+0.5A+0.5B+0.4AB,1),
\]
where $A$, $B$, and $AB$ take values $\pm1$ depending on the combination of
factor levels. For the $2^{4}$ experiment with four two-level factors, $A$,
$B$, $C$, and $D$, we use the model
\[
y_{ij}\sim N(5+0.3A+0.3B+0.3D+0.25BD,1),
\]
where $A$, $B$, $D$, and $BD$ take values $\pm1$ depending on the combination
of factor levels. We test the significance of the factorial effects of
interest for each model at the 5\% level based on the above methods.
The simulation is repeated $N=20,000$ times for each model. We compute the
percentage of rejection of the null hypothesis $H_{0}:\alpha_{l}=0$,
$l=1,\ldots,I$. Throughout this section, $I=7$ and 15 for the $2^{3}$ and $2^{4}$
designs, respectively. The results are summarized in Tables 2 and 3.

{\renewcommand{\arraystretch}{0.6} \begin{table}[!htt]
\caption{Percentage of rejection of the null hypothesis $H_{0}:\alpha_{l}=0$ at
the 5\% level for model $y_{ij}\sim N(10+ 0.5A + 0.5B + 0.4AB,1 )$ in
replicated $2^{3}$ experiments.}%
\vspace{0.01in} \centering {\tabcolsep=0.4mm
\begin{tabular}
[c]{c|cccc|cccc|cccc}\hline
Effect & $n=3$ & $n=4$ & $n=5$ & $n=6$ & $n=3$ & $n=4$ & $n=5$ & $n=6$ &
$n=3$ & $n=4$ & $n=5$ & $n=6$\\\hline
& \multicolumn{4}{c|}{Our method} & \multicolumn{4}{c|}{WH method} &
\multicolumn{4}{c}{Lenth's method}\\\hline
A & 58.7 & 75.8 & 86.0 & 92.7 & 63.3 & 77.0 & 86.4 & 92.1 & 22.9 & 27.5 &
33.1 & 37.4\\
B & 58.8 & 76.6 & 85.8 & 92.3 & 64.3 & 77.8 & 86.0 & 92.0 & 23.7 & 28.3 &
33.1 & 37.6\\
C & 4.5 & 4.6 & 4.7 & 5.3 & 5.2 & 5.1 & 4.8 & 5.3 & 0.7 & 0.5 & 0.4 & 0.5\\
AB & 41.3 & 57.0 & 68.4 & 77.3 & 45.1 & 57.8 & 69.0 & 77.8 & 13.9 & 17.6 &
21.4 & 25.2\\
AC & 4.6 & 4.8 & 5.0 & 5.4 & 5.0 & 5.0 & 4.9 & 4.8 & 0.8 & 0.5 & 0.5 & 0.4\\
BC & 4.3 & 4.8 & 4.5 & 5.4 & 4.7 & 4.9 & 5.0 & 5.1 & 0.6 & 0.5 & 0.5 & 0.4\\
ABC & 4.7 & 4.5 & 4.7 & 4.9 & 5.1 & 4.9 & 4.9 & 4.7 & 0.6 & 0.6 & 0.4 &
0.4\\\hline
\end{tabular}
}\end{table}}

{\renewcommand{\arraystretch}{0.6} \begin{table}[!htt]
\caption{Percentage of rejection of the null hypothesis $H_{0}:\alpha_{l}=0$ at
the 5\% level for model $y_{ij}\sim N(5+ 0.3A + 0.3B + 0.3D+0.25BD,1 )$ in
replicated $2^{4}$ experiments.}%
\vspace{0.01in} \centering {\tabcolsep=0.5mm
\begin{tabular}
[c]{c|cccc|cccc|cccc}\hline
Effect & $n=3$ & $n=4$ & $n=5$ & $n=6$ & $n=3$ & $n=4$ & $n=5$ & $n=6$ &
$n=3$ & $n=4$ & $n=5$ & $n=6$\\\hline
& \multicolumn{4}{c|}{Our method} & \multicolumn{4}{c|}{WH method} &
\multicolumn{4}{c}{Lenth's method}\\\hline
A & 51.2 & 63.8 & 74.1 & 81.7 & 52.6 & 65.2 & 75.2 & 82.8 & 26.2 & 33.0 &
40.7 & 48.1\\
B & 49.6 & 63.4 & 73.6 & 81.2 & 51.3 & 64.9 & 74.8 & 82.9 & 26.5 & 32.1 &
41.4 & 47.4\\
C & 4.8 & 4.6 & 4.5 & 4.5 & 5.1 & 5.0 & 4.8 & 4.7 & 1.9 & 1.6 & 1.7 & 1.8\\
D & 51.0 & 63.6 & 74.2 & 81.9 & 52.6 & 65.3 & 75.2 & 83.4 & 25.5 & 32.8 &
41.1 & 47.0\\
AB & 4.6 & 4.5 & 4.4 & 4.4 & 5.0 & 4.8 & 4.8 & 5.0 & 1.7 & 1.8 & 1.7 & 2.0\\
AC & 4.8 & 4.6 & 4.5 & 4.5 & 5.2 & 5.0 & 4.7 & 5.0 & 1.8 & 1.7 & 1.7 & 1.8\\
AD & 4.5 & 4.5 & 4.6 & 4.5 & 4.9 & 5.1 & 5.0 & 4.8 & 1.8 & 1.6 & 1.8 & 1.8\\
BC & 4.8 & 4.6 & 5.1 & 4.7 & 5.1 & 5.1 & 5.4 & 4.9 & 1.7 & 1.6 & 1.7 & 1.7\\
BD & 37.7 & 47.9 & 58.3 & 65.8 & 39.1 & 49.6 & 59.4 & 67.5 & 18.7 & 22.5 &
28.7 & 34.2\\
CD & 4.5 & 4.8 & 4.5 & 4.5 & 4.8 & 5.2 & 4.9 & 5.1 & 1.6 & 1.7 & 2.0 & 1.7\\
ABC & 5.0 & 4.5 & 4.6 & 4.5 & 5.4 & 4.9 & 4.9 & 5.1 & 1.9 & 1.7 & 1.7 & 1.9\\
ABD & 4.6 & 4.5 & 4.5 & 4.4 & 4.8 & 4.9 & 4.6 & 4.5 & 1.7 & 1.5 & 1.7 & 1.7\\
ACD & 4.6 & 4.6 & 4.5 & 4.6 & 4.6 & 4.9 & 4.7 & 5.1 & 1.8 & 1.8 & 1.5 & 1.8\\
BCD & 4.6 & 4.7 & 4.8 & 4.5 & 5.1 & 5.1 & 5.3 & 4.4 & 1.7 & 1.6 & 1.8 & 1.6\\
ABCD & 4.7 & 4.5 & 4.7 & 5.0 & 5.1 & 4.8 & 4.9 & 4.8 & 2.1 & 1.7 & 1.7 &
1.8\\\hline
\end{tabular}
}\end{table}}

From the results in Tables 2 and 3, we observe that both our
method and the WH method can tightly control the IER at the 5\% nominal level
when the $\sigma_{i}^{2}$'s are homogeneous. However, Lenth's method is unable to
tightly control the IER. In terms of power, our method has almost the same
power as the WH method in every case except for the $2^{3}$ experiment
with $n=3$. In that situation, our method is slightly less powerful.

\noindent\textbf{Case II: } ${\sigma_{i}^{2}}$ \textbf{heterogeneous}

In this case, we use the model
\[
y_{ij}\sim N\Big(10+A+B+0.5AB,\exp(A+C+0.5AC)\Big)
\]
for the $2^{3}$ factorial experiment with three two-level factors $A$, $B$, and
$C$. For the $2^{4}$ experiment with four two-level factors $A$, $B$, $C$, and
$D$, we use the model
\[
y_{ij}\sim N\Big(10+0.5A+0.45B+0.5D+0.4AD,\exp(A+B+D+0.5AD)\Big).
\]
We also test the significance of the $I$ factorial effects of interest at the
5\% level based on the above methods. For $l=1,\ldots,I$, the
percentage of rejection of the null hypothesis $H_{0}:\alpha_{l}=0$ at the 5\%
level by each method is calculated based on $N=20,000$ repetitions. The
results are summarized in Tables 4 and 5.

{\renewcommand{\arraystretch}{0.6} \begin{table}[!ht]
\caption{Percentage of rejection of the null hypothesis $H_{0}:\alpha_{l}=0$ at
the 5\% level for model $y_{ij}\sim N\big(10+ A + B + 0.5AB,\exp(A+C+0.5AC)
\big)$ in replicated $2^{3}$ experiment.}%
\vspace{0.01in} \centering {\tabcolsep=0.5mm
\begin{tabular}
[c]{c|cccc|cccc|cccc}\hline
Effect & $n=3$ & $n=4$ & $n=5$ & $n=6$ & $n=3$ & $n=4$ & $n=5$ & $n=6$ &
$n=3$ & $n=4$ & $n=5$ & $n=6$\\\hline
& \multicolumn{4}{c|}{Our method} & \multicolumn{4}{c|}{WH method} &
\multicolumn{4}{c}{Lenth's method}\\\hline
A & 57.5 & 74.3 & 86.6 & 93.1 & 71.0 & 82.3 & 90.5 & 94.8 & 31.2 & 38.6 &
44.4 & 49.9\\
B & 57.0 & 74.1 & 87.0 & 93.3 & 71.3 & 82.2 & 90.3 & 94.9 & 27.2 & 35.1 &
41.4 & 47.4\\
C & 4.7 & 4.6 & 5.2 & 5.3 & 8.3 & 7.4 & 7.1 & 7.0 & 0.2 & 0.3 & 0.2 & 0.3\\
AB & 18.3 & 24.6 & 33.8 & 41.6 & 28.5 & 33.3 & 40.6 & 46.2 & 9.8 & 12.4 &
16.3 & 20.1\\
AC & 4.5 & 4.6 & 5.1 & 5.3 & 8.5 & 7.4 & 7.0 & 6.9 & 0.3 & 0.2 & 0.2 & 0.3\\
BC & 4.6 & 4.5 & 5.3 & 5.2 & 8.2 & 7.1 & 7.4 & 6.9 & 0.5 & 0.4 & 0.3 & 0.4\\
ABC & 4.7 & 4.6 & 5.1 & 5.0 & 8.5 & 6.6 & 7.0 & 6.4 & 0.5 & 0.4 & 0.4 &
0.4\\\hline
\end{tabular}
}\end{table}}

{\renewcommand{\arraystretch}{0.6} \begin{table}[!htt]
\caption{Percentage of rejection of the null hypothesis $H_{0}:\alpha_{l}=0$ at
the 5\% level for model $y_{ij}\sim N\big(10+0.5A + 0.45B + 0.5D+0.4AD,\exp
(A+B+D+0.5AD) \big)
$ in replicated $2^{4}$ experiment.}%
\vspace{0.01in} \centering {\tabcolsep=0.5mm
\begin{tabular}
[c]{c|cccc|cccc|cccc}\hline
Effect & $n=3$ & $n=4$ & $n=5$ & $n=6$ & $n=3$ & $n=4$ & $n=5$ & $n=6$ &
$n=3$ & $n=4$ & $n=5$ & $n=6$\\\hline
& \multicolumn{4}{c|}{Our method} & \multicolumn{4}{c|}{WH method} &
\multicolumn{4}{c}{Lenth's method}\\\hline
A & 25.7 & 34.3 & 43.3 & 51.2 & 35.1 & 42.1 & 49.9 & 56.5 & 18.4 & 22.7 &
27.3 & 31.4\\
B & 21.6 & 29.4 & 36.8 & 44.2 & 30.5 & 36.7 & 43.4 & 49.1 & 16.3 & 21.0 &
25.1 & 29.0\\
C & 5.0 & 5.5 & 5.0 & 5.1 & 7.7 & 7.7 & 6.6 & 6.5 & 0.5 & 0.4 & 0.5 & 0.6\\
D & 25.7 & 34.3 & 43.0 & 51.2 & 35.5 & 42.3 & 49.8 & 56.5 & 18.5 & 23.2 &
27.0 & 31.4\\
AB & 5.1 & 5.2 & 5.2 & 4.9 & 8.0 & 7.5 & 7.0 & 6.2 & 0.5 & 0.5 & 0.4 & 0.6\\
AC & 5.1 & 5.3 & 5.2 & 5.2 & 7.8 & 7.5 & 7.0 & 6.5 & 0.6 & 0.4 & 0.5 & 0.6\\
AD & 18.6 & 24.3 & 29.9 & 36.2 & 26.1 & 30.9 & 35.7 & 41.0 & 13.5 & 16.8 &
19.9 & 22.7\\
BC & 4.9 & 5.6 & 5.1 & 5.0 & 8.0 & 7.8 & 6.7 & 6.4 & 0.6 & 0.6 & 0.5 & 0.4\\
BD & 4.9 & 4.9 & 5.1 & 5.1 & 8.0 & 7.4 & 6.8 & 6.4 & 0.6 & 0.7 & 0.5 & 0.4\\
CD & 4.9 & 5.2 & 4.9 & 5.4 & 7.9 & 7.4 & 6.6 & 6.8 & 0.5 & 0.5 & 0.5 & 0.5\\
ABC & 5.2 & 5.5 & 5.4 & 4.9 & 8.4 & 7.4 & 7.1 & 6.3 & 0.5 & 0.5 & 0.5 & 0.5\\
ABD & 5.0 & 5.1 & 5.1 & 5.0 & 8.0 & 7.6 & 6.8 & 6.2 & 0.7 & 0.5 & 0.5 & 0.5\\
ACD & 5.2 & 5.2 & 5.1 & 5.3 & 8.5 & 7.4 & 6.8 & 6.9 & 0.7 & 0.5 & 0.5 & 0.4\\
BCD & 5.1 & 5.4 & 4.8 & 5.1 & 8.1 & 7.6 & 6.7 & 6.3 & 0.7 & 0.5 & 0.5 & 0.5\\
ABCD & 5.1 & 5.3 & 5.1 & 5.0 & 7.9 & 7.6 & 6.9 & 6.4 & 0.7 & 0.5 & 0.5 &
0.5\\\hline
\end{tabular}
}\end{table}}

From the simulated results in Tables 4 and 5, we see that only our
method can tightly control the IER in all the cases for both models. These results
support our argument that the $t$-distribution with $m(n-1)$ degrees of freedom
suggested by Wu and Hamada \cite{Wu00, Wu09} may not be accurate and may fail
to control the IER when the $\sigma_{i}^{2}$'s are not the same. Again, Lenth's
method is unable to accurately control the IER in the location model.

We now compare the performance of our method, the WH method, and Lenth's
method for controlling the EER in the location model. We still consider two
cases: homogeneous ${\sigma_{i}^{2}}$'s and heterogeneous ${\sigma_{i}^{2}}$'s.

\noindent\textbf{Case I: } ${\sigma_{i}^{2}}$ \textbf{homogeneous}

Here, we consider $2^{3}$ and $2^{4}$ factorial experiments. We use the
model
\[
y_{ij}\sim N(0,1)
\]
for the simulations for both factorial experiments. The simulated EER at the
5\% level by each method is calculated based on $N=20,000$ repetitions. The
results are shown in Table 6. The table shows that all three methods
can accurately control the EER at the 5\% level.

{\renewcommand{\arraystretch}{0.6} \begin{table}[!htt]
\caption{Percentage of rejection of the null hypothesis $H_{0}:\alpha_{1}%
=\ldots=\alpha_{I}=0$ at the 5\% level for model $y_{ij}\sim N(0,1)$ in
replicated $2^{3}$ and $2^{4}$ experiments.}%
\vspace{0.01in} \centering {\tabcolsep=0.5mm
\begin{tabular}
[c]{c|cccc|cccc|cccc}\hline
I & $n=3$ & $n=4$ & $n=5$ & $n=6$ & $n=3$ & $n=4$ & $n=5$ & $n=6$ & $n=3$ &
$n=4$ & $n=5$ & $n=6$\\\hline
& \multicolumn{4}{c|}{Our method} & \multicolumn{4}{c|}{WH method} &
\multicolumn{4}{c}{Lenth's method}\\\hline
7 & 4.5 & 4.6 & 4.6 & 4.8 & 5.3 & 5.1 & 4.4 & 4.9 & 5.3 & 5.0 & 5.1 & 5.0\\
15 & 4.6 & 4.5 & 4.7 & 5.1 & 4.9 & 4.8 & 4.5 & 5.4 & 4.9 & 5.0 & 5.2 &
5.0\\\hline
\end{tabular}
}\end{table}}

\noindent\textbf{Case II: } ${\sigma_{i}^{2}}$ \textbf{heterogeneous}

Here, we use the models
\[
y_{ij}\sim N\Big(0,\mbox{exp}(A+C+0.5AC)\Big)
\]
and
\[
y_{ij}\sim N\Big(0,\mbox{exp}(A+C+D+0.5CD)\Big)
\]
for the $2^{3}$ and $2^{4}$ factorial experiments, respectively. For each
method, the EER is calculated based on 20,000 repetitions. The results are
summarized in Table 7.

{\renewcommand{\arraystretch}{0.6} \begin{table}[!htt]
\caption{Percentage of rejection of the null hypothesis $H_{0}:\alpha_{1}%
=\ldots=\alpha_{I}=0$ at the 5\% level for models $y_{ij}\sim
N\big(0,\mbox{exp}(A+C+0.5AC)\big)$ and $y_{ij}\sim N\big(0,\mbox{exp}(A+C+D+0.5CD)\big)$ in
replicated $2^{3}$ and $2^{4}$ experiments.}%
\vspace{0.01in} \centering {\tabcolsep=0.5mm
\begin{tabular}
[c]{c|cccc|cccc|cccc}\hline
I & $n=3$ & $n=4$ & $n=5$ & $n=6$ & $n=3$ & $n=4$ & $n=5$ & $n=6$ & $n=3$ &
$n=4$ & $n=5$ & $n=6$\\\hline
& \multicolumn{4}{c|}{Our method} & \multicolumn{4}{c|}{WH method} &
\multicolumn{4}{c}{Lenth's method}\\\hline
7 & 5.4 & 5.2 & 5.2 & 4.9 & 8.7 & 6.8 & 6.6 & 6.3 & 3.0 & 2.7 & 3.0 & 3.0\\
15 & 5.2 & 5.0 & 4.9 & 4.9 & 8.1 & 6.6 & 6.3 & 6.0 & 2.3 & 2.3 & 2.5 &
2.6\\\hline
\end{tabular}
}\end{table}}

Table 7 shows that only our method controls the EER at the
5\% nominal level in all the cases. The results given by the WH method are quite
{anticonservative} while those of Lenth's method are quite conservative. These results
support our argument that the studentized maximum modulus distribution with
parameters $I$ and $m(n-1)$ suggested by Wu and Hamada \cite{Wu00, Wu09} may not be
accurate and may fail to control the EER when the $\sigma_{i}^{2}$'s are heterogeneous.

\subsection{Simulation Results in Dispersion Model}

We first compare the performance of our method, the WH method, the VCA method, and
Lenth's method for controlling the IER in the dispersion model.

In the simulation, we considered $2^{3}$ and $2^{4}$ factorial experiments.
For a $2^{3}$ experiment with three two-level factors $A$, $B$, and $C$, we
generate the data using the model
\[
y_{ij}\sim N\Big(0,\mbox{exp}(0.7A+0.6C+0.6BC)\Big).
\]
Since the mean of the response does not affect the procedures
mentioned above, it is set to 0 for each run. We test the
significance of the $I=2^{3}-1=7$ factorial effects of interest at the 5\%
level based on the above procedures. For $l=1,\ldots,I$, the
percentage of rejection of the null hypothesis $H_{0}:\gamma_{l}=0$ at the 5\%
level by each method is calculated based on $N=20,000$ repetitions. The
results are summarized in Table 8.

{\renewcommand{\arraystretch}{0.6} \begin{table}[!ht]
\caption{Percentage of rejection of the null hypothesis $H_{0}:\gamma_{l}=0$ at
the 5\% level for model $y_{ij}\sim N\big(0 ,\exp(0.7A + 0.6C + 0.6BC)\big)$ in
replicated $2^{3}$ experiments.}%
\vspace{0.01in} \centering {\tabcolsep=2.8mm
\begin{tabular}
[c]{c|cccc|cccc}\hline
Effect & $n=3$ & $n=4$ & $n=5$ & $n=6$ & $n=3$ & $n=4$ & $n=5$ &
$n=6$\\\hline
& \multicolumn{4}{c|}{WH method} & \multicolumn{4}{c}{Our method}\\\hline
A & 51.2 & 65.4 & 77.8 & 85.8 & 33.7 & 53.5 & 70.0 & 80.8\\
B & 12.7 & 9.8 & 8.3 & 7.8 & 5.3 & 5.1 & 4.9 & 5.1\\
C & 41.9 & 53.9 & 65.7 & 74.6 & 25.6 & 41.5 & 56.6 & 67.9\\
AB & 12.2 & 9.7 & 8.5 & 7.6 & 5.1 & 5.2 & 5.0 & 5.1\\
AC & 12.4 & 9.5 & 8.6 & 8.2 & 5.4 & 5.1 & 5.1 & 5.2\\
BC & 42.2 & 54.0 & 65.5 & 74.0 & 25.7 & 41.8 & 56.4 & 67.4\\
ABC & 12.7 & 9.8 & 8.3 & 7.6 & 5.4 & 5.1 & 5.1 & 5.0\\\hline
& \multicolumn{4}{c|}{Lenth's method} & \multicolumn{4}{c}{VCA method}\\\hline
A & 16.4 & 20.7 & 25.5 & 29.5 & 33.1 & 52.1 & 67.7 & 78.9\\
B & 2.3 & 1.3 & 0.8 & 0.6 & 7.0 & 7.0 & 6.9 & 7.0\\
C & 11.9 & 14.7 & 18.6 & 21.6 & 26.4 & 42.0 & 54.3 & 67.1\\
AB & 1.9 & 1.1 & 0.7 & 0.4 & 6.7 & 7.1 & 6.9 & 6.9\\
AC & 2.3 & 1.3 & 0.8 & 0.6 & 6.8 & 6.7 & 6.9 & 6.9\\
BC & 11.9 & 15.0 & 18.5 & 21.5 & 26.4 & 41.6 & 55.4 & 66.9\\
ABC & 2.2 & 1.4 & 0.8 & 0.5 & 7.0 & 7.0 & 6.9 & 6.7\\\hline
\end{tabular}
}\end{table}}

For a $2^{4}$ factorial experiment with four two-level factors $A$, $B$, $C$,
and $D$, we use the model
\[
y_{ij}\sim N\Big(0,\exp(0.6A+0.6B+0.6C+0.5AD)\Big).
\]
We test the significance of each of $I=2^{4}-1=15$ effects
at the  5\% level based on all four methods. The simulation is
repeated $N=20,000$ times, and the percentage of each factorial effect
declared significant at the 5\% level is recorded in Table 9.

{\renewcommand{\arraystretch}{0.6} \begin{table}[!ht]
\caption{Percentage of rejection of the null hypothesis $H_{0}:\gamma_{l}=0$ at
the 5\% level for model $y_{ij}\sim N\big(0,\exp(0.6A + 0.6B+ 0.6C + 0.5AD)\big)$ in
replicated $2^{4}$ experiments.}%
\vspace{0.01in} \centering {\tabcolsep=2.8mm
\begin{tabular}
[c]{c|cccc|cccc}\hline
Effect & $n=3$ & $n=4$ & $n=5$ & $n=6$ & $n=3$ & $n=4$ & $n=5$ &
$n=6$\\\hline
& \multicolumn{4}{c|}{WH method} & \multicolumn{4}{c}{Our method}\\\hline
A & 63.2 & 80.1 & 89.5 & 95.0 & 46.0 & 70.5 & 84.5 & 92.8\\
B & 63.6 & 79.8 & 89.7 & 95.3 & 46.2 & 70.3 & 84.7 & 93.0\\
C & 63.6 & 79.5 & 89.9 & 95.1 & 46.2 & 69.7 & 85.4 & 92.9\\
D & 12.8 & 9.6 & 8.5 & 7.7 & 5.3 & 5.0 & 5.3 & 4.9\\
AB & 12.3 & 9.8 & 8.3 & 7.6 & 5.2 & 5.0 & 5.1 & 5.0\\
AC & 12.5 & 9.8 & 8.4 & 7.3 & 5.1 & 5.1 & 5.0 & 4.8\\
AD & 51.2 & 65.8 & 78.1 & 86.0 & 33.8 & 54.3 & 70.7 & 81.3\\
BC & 12.8 & 9.5 & 8.1 & 7.8 & 5.4 & 4.8 & 4.7 & 5.1\\
BD & 12.0 & 9.9 & 8.5 & 7.7 & 4.9 & 5.2 & 4.9 & 5.2\\
CD & 12.7 & 9.7 & 8.3 & 7.5 & 5.1 & 5.2 & 5.1 & 5.0\\
ABC & 12.7 & 9.8 & 8.6 & 7.7 & 5.0 & 5.1 & 5.2 & 5.0\\
ABD & 12.6 & 9.7 & 8.4 & 7.8 & 5.4 & 5.1 & 5.1 & 5.2\\
ACD & 12.0 & 10.0 & 8.6 & 7.9 & 4.8 & 5.3 & 5.1 & 5.2\\
BCD & 12.8 & 10.0 & 8.3 & 7.8 & 5.4 & 5.3 & 5.0 & 5.1\\
ABCD & 12.6 & 9.9 & 8.3 & 7.7 & 5.1 & 5.2 & 5.0 & 5.2\\\hline
& \multicolumn{4}{c|}{Lenth's method} & \multicolumn{4}{c}{VCA method}\\\hline
A & 23.7 & 36.8 & 50.8 & 62.7 & 45.9 & 67.9 & 83.2 & 91.9\\
B & 23.5 & 36.8 & 50.9 & 62.8 & 45.7 & 68.6 & 83.6 & 91.7\\
C & 24.0 & 37.1 & 50.8 & 62.8 & 45.4 & 67.5 & 83.6 & 91.7\\
D & 1.7 & 1.6 & 1.8 & 2.1 & 6.1 & 6.3 & 6.0 & 6.1\\
AB & 1.9 & 1.6 & 1.9 & 2.0 & 6.0 & 6.1 & 5.9 & 6.0\\
AC & 1.7 & 1.5 & 1.7 & 2.1 & 6.3 & 6.3 & 6.2 & 5.9\\
AD & 16.4 & 25.7 & 37.2 & 48.6 & 33.1 & 53.6 & 68.9 & 80.9\\
BC & 1.8 & 1.7 & 1.6 & 2.0 & 6.1 & 6.0 & 5.9 & 5.9\\
BD & 1.7 & 1.6 & 1.7 & 2.1 & 6.2 & 6.2 & 6.0 & 5.8\\
CD & 1.8 & 1.5 & 1.7 & 2.1 & 6.2 & 6.2 & 6.2 & 6.1\\
ABC & 1.7 & 1.7 & 1.7 & 2.0 & 6.0 & 6.1 & 6.0 & 5.7\\
ABD & 1.7 & 1.6 & 1.7 & 2.0 & 6.3 & 6.1 & 6.1 & 6.2\\
ACD & 1.7 & 1.8 & 1.8 & 2.2 & 6.0 & 6.1 & 6.1 & 5.9\\
BCD & 1.9 & 1.6 & 1.8 & 1.9 & 6.3 & 6.2 & 5.8 & 5.9\\
ABCD & 1.8 & 1.6 & 1.8 & 2.1 & 6.1 & 6.3 & 6.0 & 5.9\\\hline
\end{tabular}
}\end{table}}

Tables 8 and 9 show that our method achieves simulated IERs for the factorial effects not in the
models that are quite close to the 5\% nominal level. The WH method
inflates the IER, especially for small $n$; it becomes better as $n$
increases. Lenth's method is quite conservative whether
$n$ is large or small, and the VCA method is {anticonservative}. The
performance is the same for all values of $n$ considered.

We emphasize that the $z$-type statistics are the same for our method and the WH
method; the methods differ in the suggested distributions
for the $z$-type statistics. The simulation results suggest
that our suggested distribution is more accurate than that suggested by Wu
and Hamada \cite{Wu00, Wu09}.

We now compare the performance of our method, the WH method, and Lenth's method
for controlling EER in the dispersion model. Since Variyath et al. \cite{Var05} does
not have a procedure for the EER, it is not included in the comparison.

In the simulation, we considered $2^{3}$ and $2^{4}$ factorial experiments.
For each experiment, the model under the null hypothesis $H_{0}:\gamma
_{1}=\ldots=\gamma_{I}=0$ is
\[
y_{ij}\sim N(0,1).
\]
We set the mean of response to 0, since it does not affect the above
three methods.
The simulated EER at the 5\% level in the dispersion model is calculated based
on $N=20,000$ repetitions. The results are presented in Table 10.

{\renewcommand{\arraystretch}{0.6} \begin{table}[!ht]
\caption{Percentage of rejection of the null hypothesis $H_{0}:\gamma_{1}%
=\ldots=\gamma_{I}=0$ at the 5\% level for model $y_{ij}\sim N(0,1)$ in
replicated $2^{3}$ and $2^{4}$ factorial experiments.}%
\vspace{0.01in} \centering {\tabcolsep=0.6mm
\begin{tabular}
[c]{c|cccc|cccc|cccc}\hline
$I$ & $n=3$ & $n=4$ & $n=5$ & $n=6$ & $n=3$ & $n=4$ & $n=5$ & $n=6$ & $n=3$ &
$n=4$ & $n=5$ & $n=6$\\\hline
& \multicolumn{4}{c|}{WH method} & \multicolumn{4}{c|}{Our method} &
\multicolumn{4}{c}{Lenth's method}\\\hline
7 & 21.6 & 14.9 & 11.9 & 10.9 & 5.5 & 5.4 & 5.4 & 5.4 & 4.5 & 4.5 & 4.7 &
4.8\\
15 & 26.4 & 17.5 & 14.0 & 11.9 & 5.5 & 5.4 & 5.3 & 5.1 & 4.3 & 4.5 & 4.3 &
4.3\\\hline
\end{tabular}
}\end{table}}

Table 10 shows that the values for the EER based on our
method are around 5\%. This is evidence that our method can
accurately control the EER in the dispersion model. The WH method gives
results that are well above the 5\% nominal level, so this method
cannot control the EER. The EER based on
Lenth's method is quite close to the nominal level, so Lenth's method
can also tightly control the EER.

\section{Real Example}

\label{exam}

We now illustrate the application of the methods
to a real data set.

\begin{example}
This example is taken from the textbook of Montgomery \cite[p.267]{Mon09}. The
experiment is a $2^{4}$ factorial design. Four experimental factors, namely
length of putt (A), type of putter (B), break of putt (C), and slope of putt
(D), were investigated, each at two levels. The primary response in this
experiment is the distance from the ball to the center of the cup after the
ball comes to rest. The experiment is replicated seven times for each run. The
purpose of this experiment is to improve the golfer's scores (putting
accuracy), i.e., to minimize the putting variability while maintaining the
distance from the ball to the center of the cup close to zero. The data,
factors, and factor levels can be found in Montgomery  \cite{Mon09}.
\end{example}

Traditionally, the half-normal plots developed by Daniel \cite{Dan59} are used to
identify the active effects in factorial experiments.

For the location model, the half-normal plot for the 15 effects of interest, shown
in Figure 1, indicates that effect A and probably B and AB are significant. If we
control the IER of each effect in the location model at 5\%, we find that
effects A and B are declared significant by the WH method and our method, while no
effect is declared significant by Lenth's method. If we control the EER in the
location model at 5\%, both our method and the WH method find that only effect
A is significant, and Lenth's method does not identify any significant effects.

\begin{figure}[!ht]
\centering{ \includegraphics[scale=0.4]{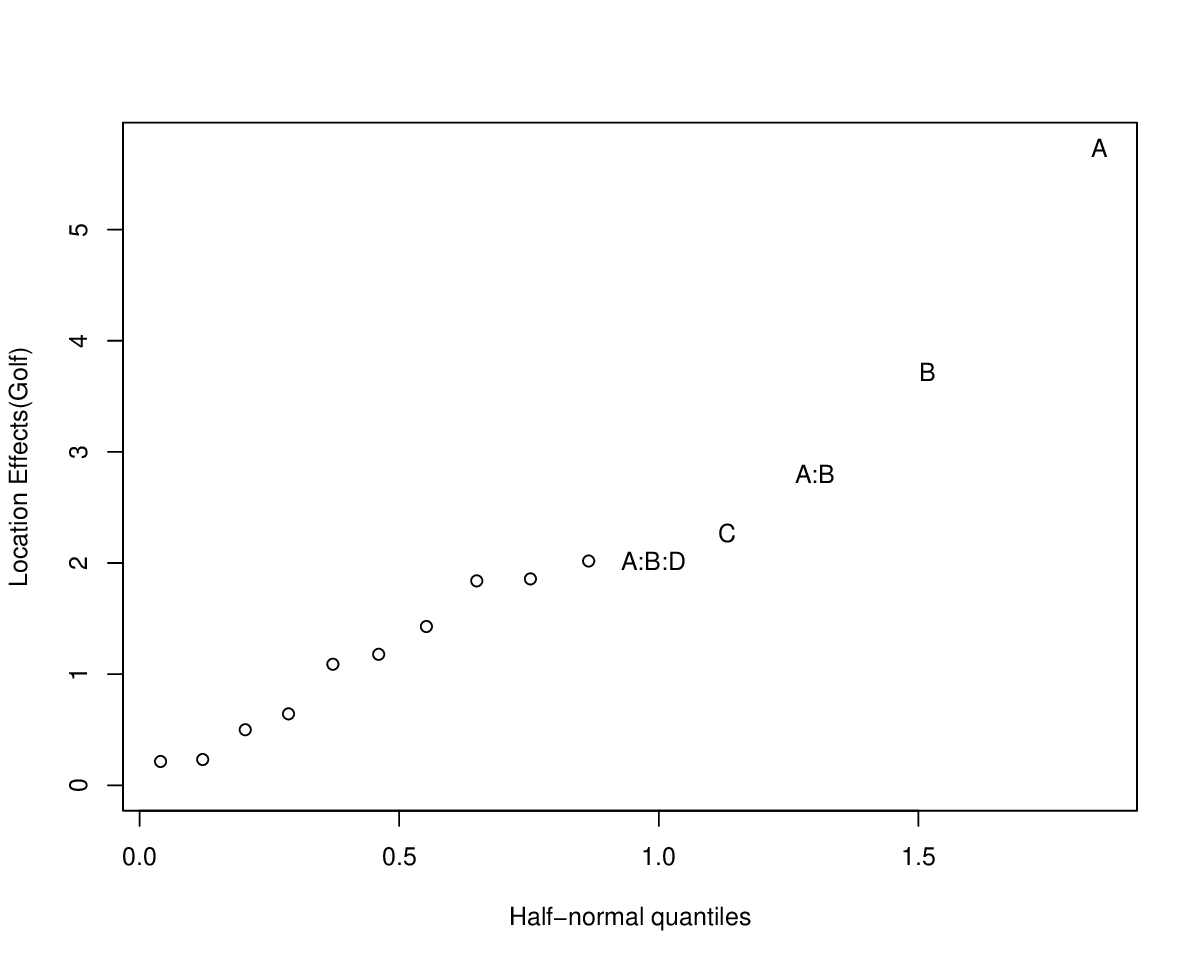} }
\caption{Half-normal plot for the 15 effects of interest in the location model.}%
\end{figure}

The half-normal plot for the 15 effects of interest in the dispersion model is
shown in Figure 2. Effect A is clearly significant, and BC, AC, ABD,
and AB are probably also significant for the dispersion model. If we control the IER of
each effect in the dispersion model at 5\%, the WH method declares that effects
A, AC, BC, and ABD are significant; both our method and the VCA method claim that
effects A and BC are significant; and Lenth's method finds that only effect A
is significant. If we control the EER in the dispersion model at 5\%, both
the WH method and our method declare that only effect A is significant, while
Lenth's method declares that no effect is significant.

\begin{figure}[!ht]
\centering{ \includegraphics[scale=0.4]{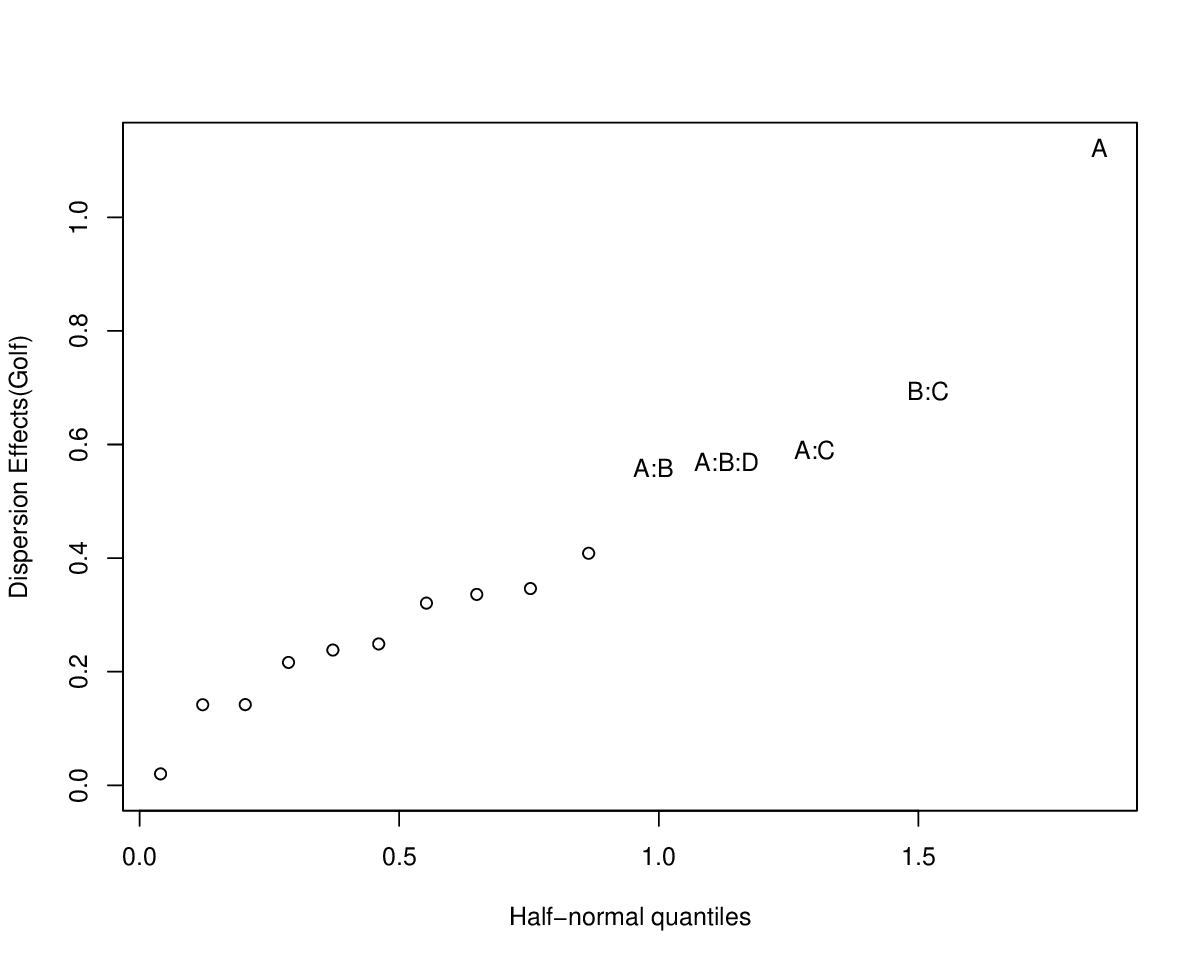} }
\caption{Half-normal plot for the 15 effects of interest in the dispersion model.}%
\end{figure}

\section{Summary}

\label{discussion}

In this paper, we have focused on controlling the IER and EER in the location and
dispersion models of the response for replicated experiments.
Specifically, our methods are based on the $t$-type test statistic \cite{Wu00, Wu09} for the factorial effects in the location model and the
$z$-type test statistic \cite{Wu00, Wu09} for the factorial effects
in the dispersion model.

We identified the true distribution of the $t$-type statistic and suggested {a Monte Carlo method}
to generate random samples from this distribution. Based on
the generated random samples, we suggested new procedures for the control of the
IER and EER in the location model. Our simulation results show that our method
works well in terms of controlling the IER and EER in the location model
whether the variances ($\sigma_{i}^{2}$'s) are homogeneous or heterogeneous over
the $m$ runs. However, the WH method works well only when the $\sigma_{i}^{2}%
$'s are homogeneous.

We re-investigated the distributions of the $z$-type statistic and
proposed a new distribution for this test statistic. Based on this
distribution, we proposed new procedures for the control of the IER and EER in the
dispersion model. Our simulation studies suggest that the new procedures work
well in terms of controlling the IER and EER in this model.
Existing methods are either {anticonservative} or conservative for the control of
either or both the IER and EER.

Identifying the active effects that affect the mean and variance of the response
is a variable selection problem in the location model (\ref{loc}) and dispersion model (\ref{dis}).
Therefore, we can apply penalized likelihood methods such as LASSO-type methods \cite{Tib96}
and SCAD-type methods \cite{Fan01}.
An interesting question is how to control the IER, EER, or false discovery rate \cite{Ben95}
in location and dispersion models
when applying a penalized likelihood method.
We leave this to future research.

\section*{Disclosure statement}


No potential conflict of interest was reported by the authors.

\section*{Funding}

This work was supported by the National Natural Science Foundation of China [grants 11771250
and 11371223], Program for Scientific Research Innovation Team in Applied Probability and
Statistics of Qufu Normal University [grant 0230518],
and the Natural Sciences and Engineering Research Council of Canada [grant RGPIN-2015-06592].

\end{document}